\documentclass[12pt]{article}
\usepackage{VMpreamble}

\usepackage{ytableau}
\usepackage{array}
\definecolor{almond}{rgb}{0.94, 0.87, 0.8}

\renewcommand{\lll}{\left\langle}
\newcommand{\rrr}{\right\rangle}

\newcommand{\ev}[1]{\Big\langle #1 \Big\rangle}

\newcommand{\QQ}{\mathbf{Q}}
\newcommand{\Tbf}{\pmb{\Theta}}
\newcommand{\ThetaW}[1]{\Theta^{\mathrm{W}}_{#1}(x|a)}
\newcommand{\HHH}{\mathbf{H}  }
\newcommand{\JJ}{\mathbf{J}^{(u,v)}}

\begin{document}

\hfill 

\hfill

\bigskip
\begin{center}
    \Large{Superintegrability of the Wilson family of matrix models and moments of  multivariable orthogonal polynomials}
\end{center}
\bigskip

\centerline{{ V. Mishnyakov}$^{a,}$\footnote{victor.mishnyakov@su.se}
	}

\bigskip

\begin{center}
	$^a$ {\small {\it Nordita, KTH Royal Institute of Technology and Stockholm University,}}\\{\small {\it 
Hannes Alfv\'ens v\"ag 12, SE-106 91 Stockholm, Sweden}}
\\
\end{center} 

\centerline{\bf \normalsize Abstract}
 \vspace{0.3cm}
        {We present new examples of superintegrable matrix/eigenvalue models. These examples arise as a result of the exploration of the relationship between the theory of superintegrability and multivariate orthogonal polynomials. The new superintegrable examples are built upon the multivariate generalizations of the Meixner-Pollaczek and Wilson polynomials and their respective measures. From the perspective of multivariate orthogonal polynomials in this work we propose expressions for (generalized) moments of the respective multivariable measures. From the perspective of superintegrability we uncover a couple of new phenomena such as the deviation from Schur polynomials as the superintegrable basis without any deformation and new combinatorial structures appearing in the answers.}

\newpage
\tableofcontents

\section{Introduction}
Matrix models are one of the central subjects in mathematical physics. They are rich enough to exhibit many properties, but also simple enough to allow for explicit computation. Recently, the property of superintegrability has been extensively discussed \cite{Mironov:2022fsr,Mironov:2020tjf,Mironov:2022gvd}. It is studied alongside the older subjects of integrability \cite{gerasimov1991matrix,morozov1994integrability,martinec1991origin}, Virasoro constraints \cite{Mironov:1990im, Cassia:2020uxy,Mironov:2021yvg} and the so-called W-representation \cite{Morozov:2009xk,Mironov:2020pcd}. The property of superintegrability (SI) is manifested as the existence of explicit formulas for expectation values when a special basis in the space of observables (symmetric functions in the case of matrix models) is chosen. A characteristic feature of SI is the rich combinatoric and representation-theoretical nature of the appearing formulas. In particular, the said basis is often given in terms of characters and their deformation such as Schur or Macdonald polynomials. At the same time the expectation values often involve the same functions, but evaluated at special points, which give rise to hook-like and content product formulas defined for integer partitions. It was recently discovered that these structures have an algebraic origin in the representation theory of certain infinite dimensional algebras, included in the family of BPS algebras \cite{Mironov:2020pcd,Mironov:2023wga}. All of the above is reminiscent of what happens in the study of localization results in gauge theories, and in this sense superintegrability of matrix models might serve as a toy model for these structures. The exact expectation values are also useful for studying other properties of matrix models. Integrability and its extension are of course the straightforward one; however, other applications involve the generalization of the enumerative geometry of the topological recursion \cite{chapuy2022non,bonzom2023b}.

In this paper, we explore the relation between the construction of multivariate orthogonal polynomials \cite{lassalle1991polynomes,dumitriu2007mops} and superintegrability of eigenvalue models.The key idea is that suitable single-variable formulas can be straightforwardly generalized to the multivariable case. In particular, this concerns formulas for moments of various standard measures. Formulas for moments utilize the so-called inversion properties for orthogonal polynomials. These are known for all polynomials in the Askey-Wilson scheme \cite{koekoek1996askey,Chan:2017qnw}, including the $q$-version. In this paper, we stay within the $q=1$ world for simplicity and leave the deformation for later. Moments of the measures of the Askey-Wilson scheme are collected in \cite{sadjang2015moments}. 

Utilizing this idea, we construct a family of new examples of matrix model superintegrability for Hermitian matrix models. Strictly speaking the resulting models are eigenvalue models, as they involve analytic continuations of contours. The constructed family of models corresponds to a multivariable generalization of the Askey-Wilson family \cite{koekoek1996askey}. From this point of view we essentially solve the moment problem for this generalization. The most general model involves the multivariate Wilson polynomials. Just as with the single variable case various limits are possible. Within these limits, most of the previously known superintegrable Hermitian matrix models are recovered.
\\

From the point of view of the theory of multivariate orthogonal polynomials this work does not present the most general notions. Still, we construct the multivariate Wilson polynomials and study of their properties. From the point of view of SI in matrix models, the new examples constructed in this work illuminates new interesting features and offers presents a new insight into the nature of this phenomenon. 

A key question in SI is the choice of basis in the space of symmetric functions (observables) that would enjoy expectation values in closed form formulas. Previously one encountered various versions and deformations of characters, such as Schur functions for the Hermitian models, Macdonald polynomials, Schur $Q$-polynomials and so on. A general expectation that one could extract from the known examples would be, that within Hermitean models we would encounter Schur polynomials independently of the potential. However, in this paper we show, that this is not necessary the case. For more complicated  potentials we find that the symmetric functions are rather different. Firstly, they are not homogeneous symmetric functions any more, and they explicitly depend on $N$ - the number of eigenvalues. Moreover, they depend on the potential and the parameters of the potential. Dependence of the symmetric function on the potential was actually already observed  for the Uglov matrix model \cite{Mishnyakov:2024cgl}. Apart from that, we try to highlight that the hypergeometric nature of the Wilson family seems to be important for superintegrability. The specific factorial-like formulas that appear within this context are naturally generalised to partitions. We will discuss this more within the paper.
\\

The paper is structured as follows.

We start with a brief reminder on the known properties of very classical orthogonal polynomials such as the Hermite and Jacobi 
and their multivariate version, including the $\beta$-deformed one. We illustrate how this is related to the strong superintegrability property explored in \cite{Mironov:2022gvd,Chan:2023lhx}

Next, as an intermediate step  towards the most general case, we consider the Meixner–Pollaczek polynomials, They are a two-parameter degeneration of the Wilson polynomials and therefore simpler.

Finally, we construct solve the multivariate moment problem for Wilson polynomials and hence present the most general superintegrability of this work.

Some notations on symmetric functions are collected in the Appendix.

\section{Multivariate very classical polynomials and superintegrability of the simplest matrix ensembles.}
We start by reviewing the relation between the theory of multivariable orthogonal polynomials and superintegrability formulas in matrix models, for two cases of the \emph{very classical} orthogonal polynomials \cite{Chan:2017qnw}: Hermite and Jacobi. These cases are well studied in the literature, see for example \cite{lassalle1991polynomes,dumitriu2007mops}
\subsection{Multivariate Hermite polynomials.}
The multivariate Hermite polynomials correspond to the Hermitean Gaussian matrix model. Here the averages are defined as
\begin{equation}
\ev{\ldots}_{\text{Gauss}} =  \int DX \exp\left(-\frac{1}{2}\Tr X^2 \right)
    \ldots
\end{equation}
Let us introduce the notion of superintegrability, using this simple matrix model as an example. In that case it is a statement about the expectation value of the Schur polynomials:
\begin{equation}
    \ev{S_R(X)}_{\text{Gauss}} = \int DX \exp\left(-\frac{1}{2}\Tr X^2 \right) S_R(X) = S_R \left\{ \delta_{k,2} \right\}\prod_{(i,j)\in R} (N+j-i)=  \dfrac{S_R\left\{ N \right\}}{S_R\left\{ \delta_{k,1} \right\}} S_R\left\{ \delta_{k,2} \right\}
\end{equation}
The notations are explained in Appendix \ref{sec:AppendixNotations}. From now on we drop the subscript, denoting the measure w.r.t which we take the expectation value, as each section is devoted to a seperate case. Since Schur polynomials form a basis in the space of invariant functions of $X$ - computation of (invariant) expectation values reduced to linear algebra, i.e. expansion of the observable in Schur polynomials. There are many ways to prove this formula. Here we would like to take an angle that relates it to multivariable Hermite polynomials and extend to later cases. 
\\\\
One motivation for that is the property that was dubbed strong superintegrability in \cite{Mironov:2022gvd}. One could ask whether superintegrability also takes place for double correlators, i.e. involving two symmetric functions. There, polynomials $K_R(X)$ where constructed, such that they enjoy special expectation values $\ev{K_R(X) S_Q(X)}_{\text{Gauss}}$ and $\ev{K_R(X) K_Q(Q)}_{\text{Gauss}}$, and are defined as:
\begin{equation}
   K_R(X) = \exp\left(\frac{1}{2}\Tr X^2 \right) S_R\left(p_k = \Tr \left( \dfrac{\partial}{\partial X} \right)^k  \right) \exp\left(-\frac{1}{2}\Tr X^2 \right) 
\end{equation}
This polynomials are essentially the multivariate Hermite polynomials, which we discuss in this section.

The multivariate Hermite polynomials are defined as the following ratio
\begin{equation}\label{eq:MultiHermite}
    \mathbf{H}_R(x_i) = \dfrac{\det_{1 \leq i,j \leq N}\left( H_{R_j+N-j}(x_i) \right)}{\Delta(x)}
\end{equation}
where $ H_{n}(x) $ are the Hermite polynomials. The divisibility of this determinant by the Vandermonde determinant follows trivially from its vanishing when any two of $x_i$ are identified. This definition makes immediate their generalizes orthogonality relation:
\begin{equation}\label{eq:MultiHermiteOrthogonal}
    \ev{ \mathbf{H}_R(x)  \mathbf{H}_Q(x)} = S_R(N) \delta_{RQ}
\end{equation}
This orthogonality under a multivariate Gaussian measure justifies the name - multivariate Hermite polynomials. At the same time, from the matrix model point of view they provide a solution to the double-correlator superintegrability. Essentially these are the polynomials of \cite{Mironov:2022gvd}, i.e. $K_R=\HHH_R$. Interestingly, these polynomials have also recently appeared in the calculation of 1/2-BPS correlators in $\mathcal{N}=4$ SYM theory \cite{Kazakov:2024ald}. To even further justify their name, let us consider some of their other properties.
\\\\
First, is their sum representation. The single variable Hermite polynomials can be written as:
\begin{equation}
    H_n(x) =  \sum_{m=0}^n 
\frac{\left(-\frac{1}{2}\right)^m n! x^{n-2
   m}}{m! (n-2 m)!}  = x^n + \ldots
\end{equation}
An analogous expansion for the multivariate case can be easily deduced if one notices that the analogue of $x^n$ in the single variable case is the Schur polynomial $S_R(x)$. Indeed the leading order term of the multivariate polynomial can be read of from its definition \eqref{eq:MultiHermite}:
\begin{equation}
    \HHH_R(x)=S_R(x) + \ldots
\end{equation}
The full expansion is given by:
\begin{equation}\label{eq:MultiHermiteExpansion}
    \HHH_R(x) = \sum_{Q \subseteq R} S_{R/Q}\left\{ \delta_{k,2} \right\} \dfrac{\xi_R(N)}{\xi_Q(N)} S_Q \,.
\end{equation}
Here we used the notation \eqref{eq:notation1}:
\begin{equation}
	\xi_R(N) =\dfrac{S_R\left\{N\right\} }{  S_R\left\{ \delta_{k,1} \right\} } =   \prod_{(i,j)\in R} (N+j-i) 
\end{equation}
Here we explicitly see the $N$-dependence of the polynomials. This expansion also has an inverse, where one expands the Schur functions as a sum of multivariate Hermite polynomials:
\begin{equation}
     S_R(x) = \sum_{Q \subseteq R} S_{R/Q}\left\{ \delta_{k,2} \right\} \dfrac{\xi_R(N)}{\xi_Q(N)} \HHH_Q
\end{equation}
which is the multivariate analogue of:
\begin{equation}
    x^n = n! \sum_{m=0}^{\left\lfloor \frac{n}{2} \right\rfloor} \frac{1}{2^m  m!(n-2m)!} H_{n-2m}(x)
\end{equation}
These so-called ''inversion'' formulas will be extremely important further. Mainly they allow to calculate the moments. We will explain the general idea later, and here just stick to this example. The orthogonality relation \eqref{eq:MultiHermiteOrthogonal} implies that $\ev{\HHH_R(x)} = \delta_{R \emptyset}$. Therefore we have:
\begin{equation}
    \ev{S_R(x)} =  \sum_{Q \subseteq R} S_{R/Q}\left\{ \delta_{k,2} \right\} \dfrac{\xi_R(N)}{\xi_Q(N)} \ev{\HHH_Q(x)} = S_{R}\left\{ \delta_{k,2} \right\} \xi_R(N)
\end{equation}
Which is exactly the superintegrability formula. Hence existence of an explicit inversion formula implies superintegrability. Note, however, that when deriving this formula, one does not use integration. The inversion also implies a stronger statement about the bilinear average of a Schur function with a multivariate Hermite polynomial, which was the main subject of \cite{Mironov:2022gvd}. Namely,
\begin{equation}
    \ev{S_R \HHH_Q} = \sum_{P \subseteq R} S_{R/P}\left\{ \delta_{k,2} \right\} \dfrac{\xi_R(N)}{\xi_P(N)} \ev{\HHH_P \HHH_Q} = S_{R/Q}\left\{ \delta_{k,2} \right\} \xi_R(N)
\end{equation}
Finally, the Hermite polynomials are often defined as solutions to a certain second order differential equation. This is how they appear, for example, in quantum mechanics. One has, for the Hermite polynomials:
\begin{equation}
    \left( \dfrac{d^2}{dx^2} + 2x \dfrac{d}{dx} +n  \right)H_n(x)=0
\end{equation}
This equation can actually be solved, by an exponential representation of Hermite polynomials or the $W$-representation:
\begin{equation}
    H_n(x) =e^{-\frac{1}{2}\frac{d^2}{dx^2} } x^n
\end{equation}
The multivariate generalization of the differential equation is given by:
\begin{equation}\label{eq:MultivarHermiteEquation}
    \left( \sum_{i=1}^N \dfrac{\partial^2}{\partial x_i^2 } -2 \sum_{i \neq j} \dfrac{1}{x_i-x_j} \dfrac{\partial}{\partial x_j} - \sum_{i=1}^N x_i \dfrac{\partial}{\partial x_i}  \right) \HHH_R(x) = -|R| \HHH_R(x)
\end{equation}
One can recognize here the rational Calogero model Hamiltonian. In this case we are actually at the free fermion point, see, however, the discussion below on $\beta$-deformation. In that way one sees that the multivariate Hermite polynomials are eigenfunctions of the rational Calogero model. This equation can be written in shorthand notation as:
\begin{equation}\label{eq:HermiteW}
   \left(W_{2} - l_0 +|R| \right)\HHH_R(x)=0
\end{equation}
Where $W_2$ denotes the operator:
\begin{equation}
    W_2= \sum_{i=1}^N \dfrac{\partial^2}{\partial x_i^2 } -2 \sum_{i \neq j} \dfrac{1}{x_i-x_j} \dfrac{\partial}{\partial x_i} 
\end{equation}
It can be rewritten in terms of power sums:
\begin{equation}
    W_2 = \sum_{n,k=1}^{\infty} k n p_{k+n-2} \dfrac{\partial^2}{\partial p_n \partial p_k}+\sum_{n,k=0}^{\infty} (n+k+2) p_{k}p_n \dfrac{\partial }{\partial p_{n+k+2}} \,\qquad p_0=N
\end{equation}
and
\begin{equation}
	l_0 = \sum_{n=1} p_n  \dfrac{\partial}{\partial p_n}
\end{equation}
The equation \eqref{eq:MultivarHermiteEquation} can be solved by an exponential operator formula or the $W$-representation:
\begin{equation}
    \HHH_R(x)=e^{\frac{W_2}{2}} \cdot S_R 
\end{equation}
Note, that this is a specific basis in the space of solutions to eq. \eqref{eq:HermiteW} defined by the property, that the leading term is a Schur polynomial. The expansion \eqref{eq:MultiHermiteExpansion} follows from the action of the $W$-operator on Schur functions:
\begin{equation}
    W_2 S_R = \sum_{Q=|R|-2} \dfrac{\xi_R(N)}{\xi_Q(N)} \lll \dfrac{\partial}{\partial p_2} S_R \big| S_Q \rrr S_Q
\end{equation}
We list these and some other properties of  the multivariate Hermite polynomials, which are analogues of the single variable case in a table in Appendix \ref{sec:AppendixHermite}.
\paragraph{Beta-deformation.}
Though deformations are not the focus of the present paper, we believe it is important, to briefly mention it here. Many properties of the multivariate Hermite polynomials can be lifted to the $\beta$-deformed case where the matrix model is substituted by the $\beta$-ensemble and Schur polynomials with Jack polynomials:
\begin{equation}
\begin{split}
    \Delta^2(x) &\longrightarrow \Delta^{2\beta}(x)
    \\
    S_R &\longrightarrow J_R
\end{split}
\end{equation}
The determinant representation is of course spoiled, but we still have the differential equation. The differential equation now is a true Calogero system, with a non-trivial coupling:
\begin{equation}
   \left( \sum_{i=1}^{N}\dfrac{\partial^2}{\partial x_i^2} - 2\beta \sum_{i \neq j}\dfrac{1}{x_i-x_j} \dfrac{\partial}{\partial x_j} -  \sum_{i=1}^{N} \dfrac{\partial }{\partial x_i} \right)  \mathbf{H}_R^{\beta}(x_i) = -|R|\mathbf{H}_R^{\beta}(x_i)
\end{equation}
We call the resulting polynomials $\beta$-multi-Hermite in this paper. The name even though clumsy, reflect their properties. These polynomials where constructed and studied within the context of the Calogero model by \cite{ujino1996algebraic,sogo1996simple} and where called the hi-Jack polynomials. 
\\

The equation also implies that the $W$-representation is preserved, since it is has the form:

\begin{equation}
    (W_{2}^{(\beta)} - l_0 ) \mathbf{H}_R^{\beta}(x_i) = -|R| \mathbf{H}_R^{\beta}(x_i)
\end{equation}
Therefore we have:
\begin{equation}
    \mathbf{H}^{\beta}(x_i)= \exp(W_{2}^{(\beta)}) J_R = \sum_{Q \subset R} \dfrac{\xi^\beta_R}{\xi^\beta_Q} J_{R/Q}\left\{ \delta_{k,2}\right\} J_Q
\end{equation}
where:
\begin{equation}
	\xi_R^\beta = \prod_{(i,j) \in R} \left( \beta N +j-1 - \beta(i-1) \right)
\end{equation}
One can invert the formula to have:
\begin{equation}
    J_R =  \sum_{Q \subset R} \dfrac{\xi^\beta_R}{\xi^\beta_Q} J_{R/Q}\left\{ \delta_{k,2}\right\}\mathbf{H}_Q^{\beta}(x_i)
\end{equation}
Hence:
\begin{equation}
    \langle J_R  \rangle = \dfrac{\xi^\beta_R}{\xi^\beta_\emptyset} J_{R/\emptyset}\left\{ \delta_{k,2}\right\} \mathbf{H}_\emptyset^{\beta}(x_i)
\end{equation}

In fact, from the differential equation one can easily prove orthogonality as well, therefore we have:
\begin{equation}
    \int \prod_{i=1}^N dx_i  \exp{\left(-\dfrac{ x_i^2 }{2}\right)} \Delta^{2\beta}(x)   \HHH_R^{\beta}(x_i)  \HHH_Q^{\beta}(x_i) \sim \delta_{RQ}
\end{equation}
\subsection{Multivariate Jacobi polynomials and the logarithmic ensemble }

Now we go on to another example of a very classical orthogonal polynomial - the Jacobi polynomial. Once again, we will describe its multivariate generalization and the related matrix model structures.
\\

Multivariate Jacobi polynomials are related to the logarithmic matrix model, which is given by the expectation value:
\begin{equation}
    \lll \ldots \rrr  = \dfrac{1}{\lll 1 \rrr}\int_{0}^{1} \prod_{i=1}^N dx_i x_i^u (1-x_i)^v  \Delta^2(x) \ldots 
\end{equation}
Superintegrability in this case is known under the name of Selberg integrals:
\begin{equation}
\ev{S_R(x)}  = \dfrac{S_R\left\{ u+N \right\} S_R\left\{ N \right\}}{S_R\left\{ u+v+2N \right\}}
\end{equation}

The single variable orthogonal polynomials, relevant for this potential are the (shifted) Jacobi polynomials:
\begin{equation}\label{eq:SingleJacobi}
   j^{(u,v)}_n(x) = \sum_{m=0}^n (-1)^{n-m} \binom{n}{m}\dfrac{(1+u+m)_{n-m}}{(1+u+v+m+n)_{n-m}}x^m
\end{equation}
Which are related to the standard textbook Jacobi polynomials by a shift of variable  and a different normalization. For  polynomials \eqref{eq:SingleJacobi} one has:
\begin{equation}
  \int_{0}^1 x^u (1-x)^v  j^{(u,v)}_n(x)  j^{(u,v)}_m(x)= \delta_{nm}\cdot \dfrac{n!(1+u)_n(1+v)_n}{(u+v+n+1)_n (u+v+2)_{2n}} \cdot\left(\int_{0}^1 x^u (1-x)^v  \right)
\end{equation}
The multivariable version is built in the same way as for the Hermite polynomials:
\begin{equation}\label{eq:MultiJacobi}
      \JJ_R(x) = \dfrac{\det\limits_{1 \leq i,j \leq N}\left( j^{(u,v)}_{R_j+N-j}(x_i) \right)}{\Delta(x)}
\end{equation}
These inherit the orthogonality relation:
\begin{equation}
    \lll  \JJ_R(x) \JJ_Q(x) \rrr =  \delta_{RQ}  \cdot ||\JJ_R||^2
\end{equation}
Where the normalization is given by:
\begin{equation}
    ||\JJ_R||^2=  \left( \prod_{i=1}^N (2 N+u+v+1-i)_{2 R_i-i+1}
   \left(N+u+v+R_i-i+1\right)_{N+R_i-i} \right)
\end{equation}
As before one has the differential equation \cite{dumitriu2007mops}, which in notation similar to \eqref{eq:HermiteW}:
\begin{equation}
   \left( W_0 + (2+u+v)l_0 + 2F_2-(u+1)F_1 \right) 
\JJ_R(x)= \left((u+v)|R| + 2\sum_{(i,j)\in R} (N+j-i) \right) \JJ_R(x)
\end{equation}
where:
\begin{equation}
\begin{split}
	&W_0 = \sum_{n,m =0}^\infty (n+m)p_n p_m \dfrac{\partial^2}{\partial p_{n+m}} + \sum_{n,m=1}^{\infty} nm p_{n+m} \dfrac{\partial^2}{\partial p_n \partial p_m}
	\\
	&F_2 =\sum_{n,m =0}^\infty (n+m+1)p_n p_m \dfrac{\partial^2}{\partial p_{n+m+1}} + \sum_{n,m=0}^{\infty} n m p_{n+m-1} \dfrac{\partial^2}{\partial p_n \partial p_m}
	\\
	&F_1 =\sum_{n=0}^\infty (n+1) p_n \dfrac{\partial}{\partial p_{n+1}}
	\end{split}
\end{equation}
We refer to \cite{Mironov:2020pcd}, for a discussion on how to rewrite various operators in power sum and $x_i$ variables, and also for explanation of the notational conventions. As with the Hermite polynomials the multivariable Jacobi polynomials can be expanded into Schur functions and vice versa:
\begin{equation}
   \begin{split}
        S_R(x) &= \sum_{Q \in R} C_{RQ} \JJ_R(x)
        \\
    \JJ_R(x)&= \sum_{Q \in R} C^{\vee}_{RQ} S_R(x)
   \end{split}
\end{equation}
The general formula for the coefficients is not known explicitly. Ways to compute them recursively are given in \cite{dumitriu2007mops} Below we present a few observations about these quantities. We will discuss the $C_{RQ}$ coefficients as they are relevant for superintegrability. Namely, they give the pair correlation functions:
\begin{equation}
    \lll S_Q \JJ_R \rrr = C_{RQ} \cdot ||\JJ_R||^2
\end{equation}
The first thing that we notice is that one can always factor out a simple term:
\begin{equation}
    C_{RQ} = 
    \dfrac{\xi_R(N)}{\xi_Q(N)} \cdot \dfrac{\xi_R(u+N)}{\xi_Q(u+N)} \cdot \tilde{c}_{RQ}(u+v)
\end{equation}
The coefficient $\tilde{c}_{QR}(u+v)$ depends only on the combination $(u+v)$ and $N$. It has now a more complicated structure, which perhaps even deviates from superintegrability. In particular, they are not always factorised. Two examples are (for $N=2)$:
\begin{equation}\label{eq:ctildeexamples}
       \begin{split}
           &\tilde{c}_{[3,2],[1]}= \frac{5 u+5 v+26}{(u+v+2) (u+v+4) (u+v+5) (u+v+6)
   (u+v+7)}
   \\ &\tilde{c}_{[6,3],[2]} =\frac{17 u^2+u (34 v+232)+v (17 v+232)+795}{(u+v+2)
   (u+v+3) \prod\limits_{i=4}^{11}(u+v+i) }
       \end{split}
\end{equation}
In the first case the numerator is linear in $u+v$, but still not of the simple form that usually appears in superintegrability formulas, while in the second example that answer is simply not factorized. These answers can be obtained from the recursive relations in \cite{dumitriu2007mops}, however, one can wonder whether there is more structure to these coefficients. We observe, that the in many cases, the non-factorizing terms can be absorbed into an evaluation of skew Schur functions at a rather special point:
\begin{equation}
    \tilde{c}_{RQ} \sim S_{R/Q}\left\{ u+v+2N + |Q|_{R}-1 \right\}
\end{equation}
where $|Q|_{R}$ is a restricted total number of boxes in $Q$:
\begin{equation}
    |Q|_{R} = \sum_{\substack{i=1 \\ Q_i \neq R_i}}^{l(Q)} Q_i
\end{equation}
For example:
\begin{equation}
    S_{[6,3]/[2]} \left\{ s \right\}=\frac{(s-1) s (s+1) (s+2) (s+3) \left(17 s^2+62
   s+60\right)}{2520}
\end{equation}
which correspond to the second example in \eqref{eq:ctildeexamples} when evaluated at $s=u+v+2\cdot 2+1$.
This observation, however, is clearly not complete, since in other cases the non-factorizing part cannot be represented in that form:
\begin{equation}
\begin{split}
        \tilde{c}_{[4,3,2],[2,1]} \Big|_{N=3}& \sim \left( 61 u^2+122 u v+990 u+61 v^2+990 v+3944 \right)
        \\
        S_{[4,3,2],[2,1]}& \sim (-120 - 94 x + 75 x^2 + 61 x^3)
\end{split}
\end{equation}
Still, we can of course recover the single Schur superintegrability property, since:
\begin{equation}
    \tilde{c}_{R\emptyset} = \dfrac{ S_R\left\{\delta_{k,1}\right\}}{ \xi_{R}(u+v+2N)}
\end{equation}
As a consequence of that we get the well known Selberg integral
\begin{equation}
    \ev{S_R(x)} = S_R\left\{\delta_{k,1}\right\}\dfrac{\xi_R(N) \xi_R(u+N)}{\xi_R(u+v+2N)}
\end{equation}

\subsection{Comment of the general structures}

Suppose we have a family of orthogonal polynomials $P_n(x)$
\begin{equation}
    \int dx \,\mu(x) P_n(x) P_m(x) = \delta_{nm}
\end{equation}
These polynomials often come with a natural expansion:
\begin{equation}\label{eq:PolynToThetaSingle}
    P_n(x)= \sum_m c_{nm} \theta_m(x)
\end{equation}
where $\theta_m(x)$ is a set of, perhaps, simpler functions, often being simply $x^m$. One can find the expectation values of $\theta_m(x)$ with the measure $\mu(x)$ if one knowns an inverse formula to the expansion \eqref{eq:PolynToThetaSingle}:
\begin{equation}\label{eq:PolynToThetaSingle}
    \theta_n(x)= \sum_m c^{\vee}_{nm} P_m(x)
\end{equation}
Then:
\begin{equation}
     \int dx \,\mu(x) \theta_n(x) = c^{\vee}_{n0}
\end{equation}
This logic was used in \cite{sadjang2015moments} to compute moments of the family of classical orthogonal polynomials. In this section we used the same arguments, but lifted to the multivariable case, to reformulate the superintegrability property of the Gaussian and Selberg models. In the next section, we utilize this point of view to find superintegrability in more involved models. We use the fact that single variable expansion formulas and their inverses are known for the Wilson family of orthogonal polynomials. As we have seen and will also see below, the hypergeometric nature of these orthogonal polynomials seems to be the reason behind the simplicity of the multivariable generalization. The general multivariable orthogonal polynomials can be defined as:
\begin{equation}
    \mathbf{P}_R(x) = \dfrac{\det\limits_{1 \leq i,j \leq N}\left( P_{R_j+N-j}(x_i) \right)}{\Delta(x)}
\end{equation}
One may hope, that if the coefficients $c_{nm}$ where simple and explicit, then the one can also generalise the expansions as follows:
\begin{equation}
    \begin{split}
        \mathbf{P}_R(x) &= 
        \sum_{Q \in R} C_{RQ} \Tbf_Q(x)
        \\
         \Tbf_R(x) &= 
        \sum_{Q \in R} C^{\vee}_{RQ} \mathbf{P}_Q(x)
    \end{split}
\end{equation}
where $ \Tbf_R(x)$ are defined in the same way:
\begin{equation}
	  \Tbf_R(x) = \dfrac{\det\limits_{1 \leq i,j \leq N}\left( \theta_{R_j+N-j}(x_i) \right)}{\Delta(x)}
\end{equation}
With the multivariate orthogonal polynomial defined in this way, orthogonality is straightforward:
\begin{equation}
    \int \prod dx_i \mu(x_i) \Delta^{2}(x_i-x_j)  \mathbf{P}_R(x) \mathbf{P}_Q(x) = \delta_{RQ}
\end{equation}
And so is the expression for the expectation value of the $\Tbf_R(x)$ polynomials.
\begin{equation}
    \int \prod dx_i \mu(x_i) \Delta^{2}(x_i-x_j)  \Tbf_R(x) = C^{\vee}_{R\emptyset}
\end{equation}
It should be noted that  by no means these statement give a way to derive the superintegrability formulas in general. Deriving the form of $C^{\vee}_{R\emptyset}$ even from the expansion of the multivariate polynomials is still a problem. With that said, we achieve two things. First, is we find another way to look at superintegrability. Second, consequently, we can try to use well known formulas for expansion \eqref{eq:PolynToThetaSingle} of some orthogonal polynomials, to give us a clue about what the special functions for superintegrability should be. The next two section are exactly an illustration of this idea.
\section{The Meixner–Pollaczek model
}

An quite instructive intermediate case in between the \emph{very classical} orthogonal polynomials and the most general Wilson polynomial are the so-called Meixner–Pollaczek polynomials. Sticking to matrix model notations, in this section we study the superintegrability and orthogonal multivariate polynomials of the matrix model:
\begin{equation}
    \int DX |\det\left( \Gamma\left( \lambda I + i X\right)\right)|^2 e^{(2\phi-\pi)\Tr X}
\end{equation}
or in eigenvalue notation:
\begin{equation}
    \int \prod_{i=1}^N dx_i \Delta^2(x) \prod_{i=1}^N \left( |\Gamma(\lambda+ i x_i)|^2\exp\left((2\phi - \pi)x_i \right) \right) 
\end{equation}
We follow the same strategy as with simpler measures. Therefore we first revisit the single variable case. In this case one has the measure:
\begin{equation}
    \mu(x)=\left( |\Gamma(\lambda+ i x)|^2\exp\left((2\phi - \pi)x \right) \right) 
\end{equation}

The orthogonal polynomials are given by the Meixner–Pollaczek (M.-P.) polynomials, typically denoted as $q_n(x|\lambda,\phi)$:
\begin{equation}
    \int dx \left( |\Gamma(\lambda+ i x)|^2\exp\left((2\phi - \pi)x \right) \right) q_n(x|\lambda,\phi) q_m(x|\lambda,\phi) = \delta_{n,m} \cdot \dfrac{2\pi \Gamma(n+2\lambda)}{(2 \sin (\phi))^{2\lambda} n!} 
\end{equation}
They are defined as:
\begin{equation}\label{eq:MPdef}
    q_n(x|\lambda,\phi)= \dfrac{(2\lambda)_n e^{in\phi}}{n!} {}_2F_1\left(  -n,\lambda+i x,2\lambda|1-e^{-2i\phi}\right) = 
\end{equation}
The integrals with this measure should be understood as a discrete sum over the poles of the $\Gamma$-function, which in the given notation for $\lambda \in \mathbb{R}$ lie on the imaginary axis. Just like  the Hermite polynomials the M.-P. polynomials enjoy several nice properties.
\\

Key to our discussion is the inversion formula, which is now somewhat different to the previous cases. Mainly the simple form is acquired not for expansion of $x^n$ into the M.-P. polynomials, but for rather for $(\lambda-i x)_n$ where:
\begin{equation}
    (x)_n =\prod_{i=1}^{n-1} (x-i)
\end{equation}
is the Pochhammer symbol. Then one has \cite{sadjang2015moments}
\begin{equation}
    (\lambda+i x)_n= \sum_{m=0}^{n} \binom{n}{m} \dfrac{(-1)^mm! (2\lambda+m)_{n-m}}{(1-e^{-2i\phi})^n e^{im\phi}} q_m(x|\lambda,\phi)
\end{equation}
This formula follows directly from the hypergeometric definition \eqref{eq:MPdef}. Therefore the moment problem is solved by:
\begin{equation}
    \ev{ (\lambda+i x)_n }  = 2 \pi  (-1)^n e^{-i \lambda  (\pi -2 \phi )} \left(1-e^{2 i
   \phi }\right)^{-2 \lambda -n} \Gamma (n+2 \lambda )
\end{equation}

According to our prescription, we can immediately generalize this to the multivariate case. In fact we do not necessarily need to generalize the M.P. polynomials to conjecture the superintegrability of this model. Instead we construct the new symmetric functions as:
\begin{equation}
    \Tbf_R(x | \lambda): =\dfrac{\det\limits_{i,j} \left( \sqrt{-1} x_i+\lambda \right)_{R_j+N-j}}{\Delta(x)}
\end{equation}
Note that these functions are not homogeneous in $x_i$ anymore. Finally, superintegrability is then given by the following formula:
\begin{equation}\label{eq:MPsuperint}
\begin{split}
       \ev{\Tbf_R(x | \lambda) }_{\mathrm{MP}}&= (-1)^{\frac{N(N+7)}{4}} S_R\left\{\delta_{k,1}\right\}   \prod_{(i,j) \in R} \left( j-i+N \right) \prod_{(i,j) \in R} \left( 2\lambda+j-i+N-1 \right) \left( \dfrac{u}{u-1} \right)^{|R|} =
        \\
        &=(-1)^{\frac{N(N+7)}{4}}\ \dfrac{ S_R\left\{N \right\}  S_R\left\{N +2\lambda-1 \right\} }{  S_R\left\{\delta_{k,1} \right\}  S_R\left\{\delta_{k,1}  \right\} }\left( \dfrac{1}{1-u} \right)^{|R|}
\end{split}
\end{equation}
where we denoted:
\begin{equation}
    u=\exp(-2 i \phi)
\end{equation}
We can notice, that, interestingly enough the answer is still expressed in terms of Schur functions, not special values of the $\Theta$-polynomials. Hence this expression solves the multivariable moment problem for the Meixner-Pollaczek measure.
\\

To justify that we have to of course provide the multivariate MP polynomials and their orthogonality themselves. Actually these where already constructed in \cite{faraut2008hermitian}. The multivariate polynomials are defined via a determinant formula (for an independent algebraic definition see \cite{faraut2008hermitian}):
\begin{equation}
\mathbf{Q}_R(\lambda,\phi,x) := \dfrac{\det
\limits_{i,j}\left( q_{R_j+N-j}(\lambda,\phi|x_i) \right) }{\Delta(x)}
\end{equation}
The orthogonality is an obvious consequence of the determinantal form, as in the general case, hence we have:
\begin{equation}
    \ev{\QQ_{R_1}(x|\lambda) \QQ_{R_2} (x|\lambda) }_{\mathrm{MP}} = \delta_{R_1,R_2} \dfrac{ S_R\left\{N \right\}  S_R\left\{N +2\lambda-1 \right\} }{  S_R\left\{\delta_{k,1} \right\}  S_R\left\{\delta_{k,1}  \right\} }\left( \dfrac{-u}{(u-1)^2} \right)^{|R|}
\end{equation}
The M.-P. polynomials have a natural expansion in terms of the Pochhammer symbols \eqref{eq:MPdef}, therefore the multivariate version has to have an expansion in terms of the $\Theta$-polynomials. This is indeed the case, the expansion goes as:
\begin{equation}
    \mathbf{Q}_R(\lambda,\phi|x)= \sum_{Q \subseteq R } (-1)^{\frac{3|R|}{2}+\frac{N(N-1)}{4}}\dfrac{\prod\limits_{(i,j) \in R}(2\lambda+N+j-i-1)(N+j-i)}{\prod\limits_{(i,j) \in Q}(2\lambda+N+j-i-1)(N+j-i)} \cdot \dfrac{S_{R/Q}\left\{ \delta_{k,1} \right\}}{(1-u)^{|R|-|Q|}} \Tbf_Q(x|\lambda)
\end{equation}

This expansion can be inverted to give:
\begin{equation}
    \Tbf_R(x)=\sum_{Q \subseteq R } (-1)^{\frac{|Q|}{2}+\frac{N(N+7)}{4}}\dfrac{\prod\limits_{(i,j) \in R}(2\lambda+N+j-i-1)(N+j-i)}{\prod\limits_{(i,j) \in Q}(2\lambda+N+j-i-1)(N+j-i)} \cdot \dfrac{S_{R/Q}\left\{ \delta_{k,1} \right\}}{(u-1)^{|R|-|Q|}}  \QQ_Q\left(\lambda,\phi|x \right)
\end{equation}
Plugging this expansion into the expectation value for $\Tbf$ and using orthogonality we immediately get the superintegrability formula \eqref{eq:MPsuperint}.
\\\\
As we can see the overall pattern of generalisation from single to multivariate case is quite transparent. Mainly, one substitutes:
\begin{equation}\label{eq:hypergeom}
\begin{split}
        (x)_n & \quad \longrightarrow \quad \prod_{(i,j) \in R}(x+j-i)\\
        x^n & \quad \longrightarrow  \quad x^{|R|}
\end{split}
\end{equation}
As we have seen the multivariate Hermite equation plays quite an important role - it relates the considered function to the theory of integrable systems. An analogous equation exists for the multivariate M.-P. polynomials. In this case, however, the equations are difference. For the single variable case we have:
\begin{equation}
    e^{-i \phi}(i x +\lambda) \left[ q_n(x)-q_n(x-i) \right] + e^{i \phi}(i x - \lambda) \left[ q_n(x) -q_n(x+i)\right] = 2n i \sin\left( \phi \right) q_n(x)
\end{equation}
While for the multivariate polynomials, we have an equation \cite{faraut2008hermitian} 
\begin{equation}
\begin{split}
        &N e^{i\phi} \sum_{j=1}^N (\lambda-i x_j) \left(\prod_{k \neq j} \dfrac{x_j-x_k+i}{x_j-x_k} \right) \left[ \mathbf{Q}_R(x_j+i) -\mathbf{Q}_R(x_j) \right]  \  -
        \\
        & \quad - N e^{-i \phi} \sum_{j=1}^N (\lambda+i x_j) \left(\prod_{k \neq j} \dfrac{x_k-x_j+i}{x_k-x_j}\right) \left[\mathbf{Q}_R(x_j-i) -\mathbf{Q}_R(x_j) \right] = 2i |R| \sin(\phi) \mathbf{Q}_R(x)
\end{split}
\end{equation}
At can be also written in differential representation:
\begin{equation}
\begin{split}
        &\left(N e^{i\phi} \sum_{j=1}^N (\lambda-i x_j) \left(\prod_{k \neq j} \dfrac{x_j-x_k+i}{x_j-x_k} \right) \left[ e^{i \partial_j} -1 \right] \right.  - 
        \\
        & \qquad \qquad \left. - N e^{-i \phi} \sum_{j=1}^N (\lambda+i x_j) \left(\prod_{k \neq j} \dfrac{x_k-x_j+i}{x_k-x_j} \right) \left[e^{-i \partial_j} -1 \right] \right) \mathbf{Q}_R(x) = 2i |R| \sin(\phi) \mathbf{Q}_R(x)
\end{split}
\end{equation}
It's relation to an integrable system and the corresponding higher Hamiltonians are to be found in the future. Due to the difference nature of the equations it is likely related to some kind of q-deformed Hamiltonian, but evaluated at special values of $q$. 

\section{The Wilson model}

Finally, in this section we present inversion formulas and the SI of the multivariable case of Wilson polynomials.Let briefly collect some facts about the single variables Wilson polynomials once again. They are defined as:
\begin{equation}
   \dfrac{W_n(x|a,b,c,d)}{(a+b)_n (a+c)_n (a+d)_n }={}_4 F_3 \left( \begin{gathered}
        -n,n+a+b+c+d-1,a+ix, a-i x\\
        a+b,a+c,a+d  
    \end{gathered}
     \  \Bigg|\  1 \right)
\end{equation}
which means that they are expanded nicely into the following combination of Pochhammer symbols:
\begin{equation}
    \theta_n(x|a)=(a-ix)_n(a+i x)_n
\end{equation}
These polynomials are orthogonal w.r.t to the measure:
\begin{equation}
    w(x)=\left|
{\Gamma(a+ix)\Gamma(b+ix)\Gamma(c+ix)\Gamma(d+ix)\over\Gamma(2ix)}\right|^2
\end{equation}
hence:
\begin{equation}
    \int_{0}^{\infty} W_n(x|a,b,c,d) W_m(x|a,b,c,d)= n!\frac{\Gamma_{n+a+b} \Gamma_{n+a+c} \Gamma_{n+a+d} \Gamma_{n+b+c} \Gamma_{n+b+d} \Gamma_{n+c+d} }{\Gamma_{a+b+c+d+n-1}(a+b+c+2n-1)} \delta_{m n}
\end{equation}
where:
\begin{equation}
	\Gamma_x = \Gamma(x)
\end{equation}
The inversion formulas and hence the solution to the single variable moment problem were given in \cite{sadjang2015moments}:
\begin{equation}\label{eq:ExpansionSingleWilson1}
    \theta_n(x|a)=\sum_{m=0}^n\binom{n}{m} \frac{(-1)^m(a+b+m)_{n-m}(a+c+m)_{n-m}(a+d+m)_{n-m}}{(a+b+c+d+m-1)_m(a+b+c+d+2 m)_{n-m}} W_m\left(x | a, b, c, d\right)
\end{equation}
and, hence, the moment are given by:
\begin{equation}
    \ev{\theta_n(x|a)} = 2\pi \dfrac{\Gamma_{a+b}\Gamma_{a+c}\Gamma_{b+c}\Gamma_{b+d}\Gamma_{c+d}}{\Gamma_{a+b+c+d}} \cdot \dfrac{(a+b)_n (a+c)_n (a+d)_n}{(a+b+c+d)_n}
\end{equation}
As we can see the formulas are mostly made out from Gamma functions again, which allows us the guess the multivariable generalisation.
\\

First construct the multivariable $\Theta$ and Wilson polynomials:
\begin{equation}
\begin{split}
    \Theta^{\mathrm{W}}_R(x|a)  : =\dfrac{\det\limits_{i,j} \theta(x_i|a)_{R_j+N-j}}{\Delta(x)}
    \\ \tilde{\mathbf{W}}_R(x|a)  : =\dfrac{\det\limits_{i,j} W_{R_j+N-j}(x_i|a)}{\Delta(x)}
\end{split}
\end{equation}
In the case of Wilson polynomials the expansion formulas turn out to be more complicated. This is related to the non-standard structure of the denominator in the single variable formula \eqref{eq:ExpansionSingleWilson1}: $(a+b+c+d+2 m)_{n-m}$. The slight difference from the standard structure \eqref{eq:hypergeom} make the multivariable deformation not evident. We will discuss this issue below. First, it makes sense to absorb the second factor in the denominator in \eqref{eq:ExpansionSingleWilson1} in the normalisation, to make the expansion simpler. Hence we first redefine the normalised multivariable Wilson polynomials to be:
\begin{equation}
    \mathbf{W}_R(x|a)  : =\dfrac{\det\limits_{i,j} \left(\dfrac{ W_{R_j+N-j}(x_i|a)}{(a+b+c+d+R_j+N-j-1)_{R_j+N-j}} \right) }{\Delta(x)}
\end{equation}
The expansion formulas have the general form as usual:
\begin{equation}\label{eq:ExpansionMultiWilson}
\begin{split}
          \Theta^{\mathrm{W}}_R(x|a) = \sum_{Q \subseteq R} C_{RQ}(a,b,c,d)\mathbf{W}_Q(x|a)
          \\ 
           \mathbf{W}_R(x|a)  =  \sum_{Q \subseteq R}  C^{\vee}_{RQ}(a,b,c,d)\Theta^{\mathrm{W}}_Q(x|a)
\end{split}
\end{equation}
With the normalised definition of the Wilson polynomials we have:
\begin{equation}
    C_{RR}=C^{\vee}_{RR}=1 \,.
\end{equation}
The numerator in the single variable expansion has the standard structure that can be generalised to the multivariate case. Therefore we claim:
\begin{equation}
    C_{RQ} = \dfrac{\xi_{R/Q}(N) \xi_{R/Q}(N+a+b)\xi_{R/Q}(N+a+c)\xi_{R/Q}(N+a+d)}{\alpha^{W}_{R,Q}(a,b,c,d)}
\end{equation}
where we have introduced a shorthand notation:
\begin{equation}
    \xi_{R,Q}(u) = \dfrac{\prod\limits_{(i,j)\in R}(u+i-j)}{\prod\limits_{(i,j)\in Q}(u+i-j)}
\end{equation}
Note that this quantity is not given by a evaluation of the skew-Schur polynomial:
\begin{equation}
    \xi_{R,Q}(u)=\dfrac{S_R\left\{ p_k=u \right\}}{S_Q\left\{ p_k=u \right\}} \neq S_{R/Q}\left\{ p_k=u \right\}
\end{equation}
By $\alpha^{W}_{R/Q}(a,b,c,d)$ we denote a more complicated, and yet unknown denominator structure. As we have not been above to come up with a general formula for this denominator, we will simply present a few observations about this quantity. 
\begin{itemize}
    \item For single row partitions one has: 
    \begin{equation}
       \alpha^{W}_{[r],[q]}(a,b,c,d) = (a+b+c+d+q-1+2N)_{r-q}  
    \end{equation}
    For $N=1$ this is the single variable case. Since these numerators will always depend on $a+b+c+d$ we denote this sum by:
    \begin{equation}
        z=a+b+c+d
    \end{equation}
    \item For two-rowed partitions we were able to observe the following formula:
    \begin{equation}
       \hspace{-1.1cm} \alpha_{[r_1,r_2],[q_1,q_2]}= \left(2 N+z+2 q_1-2\right)_{r_1-q_1}  \cdot \left\{ \begin{split}
            & \frac{\left(2 N+z+2
            	q_2-4\right)_{\min \left(r_1,q_1-q_2+r_2\right)-q_2+1}}{2
            	N+q_1+q_2+z-4} \, , \quad q_1>r_2\\
            &\left(2 N+z+2
            q_2-4\right)_{r_2-q_2} \, , \quad q_1<r_2
        \end{split} \right.
    \end{equation}
    \item For three-row partitions we provide a few examples with $N=3$:
    \begin{equation}
    \begin{split}
                &\alpha_{[3,2,1],[2,1]} = 
                \quad \alpha_{[3,3,2],[3,2,1]} = (z+2)(z+6)
                  \\
                & \alpha_{[3,3,1],[3,1]} = z(z+4)(z+5)
                 \quad \alpha_{[3,3,1],[3,1]} = z(z+4)(z+5)
                \\
                &\alpha_{[3,2,1],[2]} = (z+1)(z+2)(z+3)(z+8)
                  \quad \alpha_{[2,2,2],[1,1,1]} = (z+4)(z+5)(z+6)
                   \\
                &\alpha_{[3,3,3],[2,2,1]} = (z+2) (z+5) (z+7) (z+8)
    \end{split}
    \end{equation}
\end{itemize}
We can clearly observe, that the structure of the coefficients depends in some way on whether the skew diagram is disconnected or not. In other words it depends not only on contents of individual boxes, bot also on whether they are connected within a hook shape in the skew partition. When the pieces of the skew partition are not ''interacting'' the formula for the coefficients is quite simple:
\begin{equation}
    \alpha^{\mathrm{W}}_{RQ}=\prod_{i=1}^{l(R)}  (z+2(N-1+Q_i) )_{R_i-Q_i} \, \quad Q_{i}<R_{i+1} \forall i
\end{equation}
and reflects this non-interaction by having boxes in each row depend only on the data of the same row. However, when the rows of the skew partitions overlap - they interact and the formula becomes more complicated. We were not able to deduce it here, except for the length 2 partitions, and leave this for future work.
\\\\
With the given definition of the normalized Wilson polynomials, orthogonality of the multivariate polynomials takes the form:
\begin{equation}\label{eq:OrthogonalityMultiWilson}
    \int \prod_{i=1}^N dx_i w(x_i) \Delta^2(x)     \mathbf{W}_R(x|a)   \mathbf{W}_Q(x|a) = \delta_{RQ} \cdot 
\end{equation}
where each integral over $x_i$ is understood as a contour integral like in the single variable case. Finally, using the expansion formulas \eqref{eq:ExpansionMultiWilson} and orthogonality \eqref{eq:OrthogonalityMultiWilson} we get the statement of superintegrability of the Wilson distribution or:
\begin{equation}
     \int \prod_{i=1}^N dx_i w(x_i) \Delta^2(x) \ThetaW{R} = S_R\left\{N \right\} \dfrac{\xi_R(N+a+b)\xi_R(N+a+c)\xi_R(N+a+d)}{\xi_R(2(N-1)+a+b+c+d)}
\end{equation}
To conclude this section we note that the single variable Wilson polynomials also satisfy a difference equation:
\begin{equation}
    B(x)\left(W_n(x+i)-W_n(x)\right)+W_n(x)+\left(W_n(x-i)-W_n(x) \right) =n(n-1+a+b+c+d)W_n(x)
\end{equation}
where:
\begin{equation}
    B(x)=\frac{(a-i x) (b-i x) (c-i
   x) (d-i x)}{2ix (2 i x-1)} \qquad D(x)=\frac{(a+i x) (b+i x) (c+i
   x) (d+i x)}{2ix (2 i x+1)}
\end{equation}
As with the M.-P. case, it would be interesting to come up with multivariable difference equation and study its integrable nature.
\section{Discussion and conclusions}
We have explored the relation between the superintegrability property of  matrix models and the theory of multivariable orthogonal polynomials. From the latter point of view, we addressed the multivariable version of calculating the moments of the respective measures. As a part of that we have also introduced a working definition of full multivariable Wilson polynomials, and listed some of their properties. From the SI perspective, we have enriched the class of superintegrable models with a class of new examples within the hermitian eigenvalue models. These new models have quite different potentials from the ones considered previously. The key feature is that the type of symmetric functions that allow for explicit expectation values are different from Schur functions, even though we are within the Hermitian case. 

As we have noted throughout the paper, the simple generalization from single to multivariable case seems to be related to the hypergeometric nature of the Wilson family of orthogonal polynomials. All the encountered coefficients are of factorial nature and get substituted by content products over partitions. It would be interesting to investigate this property further.

Questions of $\beta$ and $(q,t)$-deformation of these new SI examples are beyond the scope of this work, however, are very interesting. Perhaps a more invariant algebraic approach, similar to the one taken in \cite{faraut2008hermitian} should reveal, what is the correct $\beta$-deformation of the $\Theta$-polynomials and which measure they would correspond to.

Finally, the whole discussion has to be extended to discrete and $q$-orthogonal polynomials. The analogue formulas should be kind of dual to the ones considered here and should involve sums over partitions like for matrix model (or generalized) partition functions and more importantly supersymmetric gauge theory partition functions. 

\paragraph{Acknowledgements.}
\noindent We thank T.~Kimura, A.~Morozov, A.~Mironov, Y.~Zenkevich, N.~Tselousov, A.~Popolitov and V.~Kazakov for illuminating discussions. Nordita is supported in part by NordForsk.

\bibliographystyle{utphys.bst}
\bibliography{MOPSandSI,Uglov}

\appendix 
\section{Symmetric functions and notations}\label{sec:AppendixNotations}
We briefly introduce a few notations from the theory of symmetric functions and partitions that are used throughout the paper \cite{Mironov:2022fsr,macdonald1998symmetric}. 
\\

Integer partitions and Young diagrams are denoted by $R  = [R_1,R_2,R_3, \ldots]$. Superintegrability formulas are often given in terms of the so-called contents of the partitions. It is defined in terms of the coordinates of a box in the Young diagram $(i,j)$ with $i$ being the row number and $j$ the column number.  For example, the shaded box in the figure below has coordinates $(2,3)$:
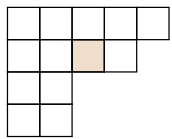
\vspace{0.3cm}
\begin{center}
    \ytableausetup{boxsize=1em,aligntableaux = center}  \quad \ydiagram{5,4,2,2}*[*(almond) ]{0,2+1}
\end{center}
\vspace{0.3cm}
The difference $j-i$ is called a content of a given box of a partition. Next, we illustrate the notations for special functions evaluated at special points on the example of Schur functions. Schur functions are characters of $GL(N)$ and can be computed in several ways. In particular, consider a generating function:
\begin{equation}
    \exp\left( \dfrac{z^k p_k}{k} \right) = \sum_k s_k(p) z^k
\end{equation}
The Schur polynomials are given by the determinant:
\begin{equation}
    S_R(p_k) = \det_{i,j} s_{R_i-i+j}(p_k)
\end{equation}
In this form, Schur functions are homogeneous functions of the variables $p_k$ of degree $R$, if $p_k$ is assigned a degree $k$. In symmetric functions notations $p_k$ are nothing but the powers sums:
\begin{equation}\label{eq:powersum}
    p_k = \sum_{i=1}^N x_i^k
\end{equation}
Making this substitution in Schur functions turns them into a symmetric functions of the $x_i$ variables. These can also be though as being eigenvalues of some matrix $X$, then one has:
\begin{equation}
    S_R(x_i)= S_R(p_k = \Tr X^k)
\end{equation}
In matrix models this is exactly how the Schur function appears in the integrand. On the r.h.s of superintegrability formulas we encounter Schur functions evaluated at special loci. Everywhere we mean special loci of the $p_k$ variables. For that we use a special notation, for example:
\begin{equation}\label{eq:not1}
    S_R\left\{ \delta_{k,s} \right\} := S_R(p_k = \delta_{k,s})
\end{equation}
which means we put all powers sums equal to zero except the $s$'th one, which is equal to one. We also use:
\begin{equation}\label{eq:not2}
     S_R\left\{ N \right\} := S_R(p_k = N )
\end{equation}
where all powers sums are equal to $N$. While the first point is natural only in the power sum basis, the second one be also expressed in the $x_i$, by putting all $x_i=1$. These special values of Schur functions can be described combinatorially as follows:
\begin{equation}
    \begin{split}
         &S_R\left\{ \delta_{k,s} \right\}   = \prod_{(i,j) \in R} \dfrac{1}{[[h_{i,j}]]_{s,0}} 
         \\
        &S_R\left\{N\right\} =  S_R\left\{ \delta_{k,1} \right\}  \prod_{(i,j)\in R} (N+j-i)
    \end{split}
\end{equation}
where $h_{i,j}$ are the corresponding hook length. i.e. the number of boxes right and below of the given one plus one, and:
\begin{equation}
	[[x]]_{s,a} = \left\{ \begin{split} &x \, , \, x=a\mod s
		\\
		&1 \, , \, x=\neq a\mod s
		\end{split}  \right.
\end{equation}
Throughout the paper we also use the notation:
\begin{equation}\label{eq:notation1}
	\xi_R(z) =\dfrac{S_R\left\{z\right\} }{  S_R\left\{ \delta_{k,1} \right\} } =   \prod_{(i,j)\in R} (z+j-i) 
\end{equation}
  With other symmetric functions that appear in the paper we use the same notation as in \eqref{eq:not1} and \eqref{eq:not2}, meaning that one expresses then in the power sum basis and evaluates at special values of the power sums.

\section{Properties of multivariate Hermite polynomials}\label{sec:AppendixHermite}
In this section we provide an incomplete list of properties of multivariate Hermite polynomials, which are counterparts of single the variable case.

\begin{table}[H]
		\vspace{0.3cm}
\begingroup
\centering
\renewcommand{\arraystretch}{2.6}
\newcolumntype{M}[1]{>{\centering\arraybackslash }m{#1}<{}}
\hspace{0cm}
	\begin{tabular}{M{0.4\linewidth}|M{0.5\linewidth}}
  Single variable & Multivariable
    \\ \hline
	    $H_n(x) \, ,  \quad n \in \mathbb{Z} $ & $\HHH_R(x) \, , \quad R \ - \ \text{partition}$ \\
        
        \multicolumn{2}{c}{Recurrence/Pieri rule }\\
\hline
    $x H_n = H_{n+1} +n H_{n-1}$ &  $p_1 \HHH_R  = \sum\limits_{R+\Box} \HHH_{R+\Box} + \sum\limits_{R-\Box} (N+j_\Box-i_\Box)\HHH_{R-\Box}$
    \\
   
     \multicolumn{2}{c}{Differentiation}\\ \hline
    $ \frac{d}{dx}H_n = n H_{n-1}$ & $\sum\limits_{i=1}^N \dfrac{\partial }{\partial x_i} \HHH_R = \sum\limits_{R-\Box} (N+j_\Box-i_\Box) \HHH_{R-\Box}$
    \\ 
     \multicolumn{2}{c}{Rodrigues formula }\\ \hline
    $H_n(x)= (-1)^n e^{x^2/2} \frac{d^n}{dx^n} e^{-x^2/2} $ & {\footnotesize$  \HHH_R = \exp\left(\frac{\Tr X^2}{2} \right) S_R\left(p_k = \Tr \left( \dfrac{\partial}{\partial X} \right)^k  \right) \exp\left(-\frac{\Tr X^2}{2} \right) $}\\
     \multicolumn{2}{c}{Differential equation}\\ \hline
    $    \left( \dfrac{d^2}{dx^2} + 2x \dfrac{d}{dx} +n  \right)H_n(x)=0$ &$   \left(W_{2} - l_0 +|R| \right)\HHH_R(x)=0$\\
    
     \multicolumn{2}{c}{Exponential/W-representation}\\
     \hline
    $ H_n(x) =e^{-\frac{1}{2}\cdot\frac{d^2}{dx^2} } x^n$ &$\HHH_R(x)=e^{\frac{W_{2}}{2}} \cdot S_R $
    
	\end{tabular}

\endgroup
\vspace{0.2cm}
\caption*{Common properties of the single and multivariable Hermite polynomials }
		\label{table:eqmassD2}
		\vspace{0.3cm}
\end{table}

Additionally, we provide two expression for multi-variate Hermite polynomials for specially shaped partitions

\begin{equation}
\begin{split}
		H_{[N-1,N-2,N-3]}(x) &= (-1)^{\frac{1}{2} (N-2) (N+1) 	}\prod_{i>j=1}^{N}(x_i+x_j)\\
			H_{[N,N-1,N-2]}(x) &=(-1)^{\frac{1}{2} (N-2) (N+1) 	} \prod_{i=1}^{N} x_i \prod_{i>j=1}^{N}(x_i+x_j)
	\end{split}
\end{equation}

\end{document}